# The impact of hydrodynamic interactions on protein folding rates depends on temperature

Fabio C. Zegarra,[1,2] Dirar Homouz,[1,2,3] Yossi Eliaz,[1,2] Andrei G. Gasic,[1,2] and Margaret S. Cheung[1,2]*

[1]*Department of Physics, University of Houston, Houston, Texas 77204, USA*

[2]*Center for Theoretical Biological Physics, Rice University, Houston, Texas 77005, USA*

[3]*Khalifa University of Science and Technology, Department of Physics, P.O. Box 127788, Abu Dhabi, United Arab Emirates*

*Corresponding author: mscheung@uh.edu

# ABSTRACT

We investigated the impact of hydrodynamic interactions (HI) on protein folding using a coarse-grained model. The extent of the impact of hydrodynamic interactions, whether it accelerates, retards, or has no effect on protein folding, has been controversial. Together with a theoretical framework of the energy landscape theory (ELT) for protein folding that describes the dynamics of the collective motion with a single reaction coordinate across a folding barrier, we compared the kinetic effects of HI on the folding rates of two protein models that use a chain of single beads with distinctive topologies: a 64-residue α/β chymotrypsin inhibitor 2 (CI2) protein, and a 57-residue β-barrel α-spectrin src-Homology 3 domain (SH3) protein. When comparing the protein folding kinetics simulated with Brownian dynamics in the presence of HI to that in the absence of HI, we find that the effect of HI on protein folding appears to have a "crossover" behavior about the folding temperature. Meaning that at a temperature greater than the folding temperature, the enhanced friction from the hydrodynamic solvents between the beads in an unfolded configuration results in lowered folding rate; conversely, at a temperature lower than the folding temperature, HI accelerates folding by the backflow of solvent toward the native folded state. Additionally, the extent of acceleration depends on the topology of a protein: for a protein like CI2, where its folding nucleus is rather diffuse in a transition state, HI channels the formation of contacts by favoring a major folding pathway in a complex free energy landscape, thus accelerating folding. For a protein like SH3, where its folding nucleus is already specific and less diffuse, HI matters less at a temperature lower than the folding temperature. Our findings provide further theoretical insight to protein folding kinetic experiments and simulations.

# I. INTRODUCTION

Aqueous solvent plays an active role in the dynamics of proteins by inducing the hydrophobic collapse of the chain and helping in the search of the specific three-dimensional structure to perform their biological function [1]. However, the motions of the solute particles of the protein are not independent and intimately coupled by the solvent. As solute particles move, they induce a flow in the solvent, which, in turn, affects the motion of neighboring solute particles. These long-range interactions between solute particles and solvent, known as hydrodynamic interactions (HI), are studied extensively in polymers both analytically [2,3] and numerically [4]. HI generally accelerates the speed of collapse [5-8] when a polymer is quenched from good to poor solvent at θ temperature. Unlike a homopolymer, a heteropolymeric protein is made up of 20 different amino acids that interact through electrostatics or van der Waals forces to various extents. These interactions are long-range in nature, which complicates the analysis of HI in protein folding. Several groups have employed computer simulations [9-12] on the investigation of HI effects on protein folding with Langevin dynamics.

Up to now the outcome from coarse-grained protein folding simulations on whether HI accelerates or deters protein folding often varies by research groups. The Cieplak group and the Elcock group showed that HI moderately accelerates the folding kinetic rates by a factor of 1.2 to 3.6 [9,10]. A recent study by the Scheraga group argued that HI reduces the folding kinetic rates [11]. Furthermore, Kikuchi *et al*. [12] claimed that HI has accelerated kinetic rates, albeit a small effect. Noticeably, there is scarce work being done on the temperature dependence of these findings. The discussion about temperature is necessary because protein folding from an unfolded configuration to a natively compact one requires imperfect cancellation of configurational entropy loss and enthalpy gain during the course of collapse, which gives rise to a temperature dependent activation barrier [13,14]. Without a comprehensive investigation over a wide range of temperature, it is challenging to delineate the real impact of HI on protein folding rates. Despite the confounding results from the groups mentioned above, they all might be correct at their own specific temperature range.

Our motivation is to reconcile the differences in reported influences of HI on protein folding over a wide range of temperature from the viewpoint of the folding energy landscape theory [14,15], particularly with a funnel-shaped energy landscape [16]. We used a computer protein model that guarantees to fold into the native state from any unfolded conformation [17]. We tracked its collective motion on a single reaction coordinate, the fraction of the native contact formation $Q$ either on a thermodynamic free energy barrier or by kinetic trajectories. We studied the effects of HI on folding of two well-studied, model proteins with distinctive topologies: one is the 64-residue $\alpha/\beta$ protein chymotrypsin inhibitor 2 (CI2) [18] shown in Fig. 1(a), and the other is the 57-residue $\beta$-barrel $\alpha$-spectrin Src-homology 3 (SH3) domain [19] shown in Fig. 1(b). The two proteins fold and unfold in a two-state manner and have been used for studying folding mechanisms from other computational studies [20-24]. We simulated the Brownian dynamics of particles including HI by implementing the algorithm developed by Ermak-McCammon [25]. The effects of HI are approximated through a configuration-dependent diffusion tensor **D** used in the Brownian equation of motion.

Our study shows that the effect of HI on folding rates can both accelerate protein folding at a temperature lower than the folding temperature and retard protein folding speed at a temperature higher than the folding temperature, in comparison with the folding dynamics without HI. Since HI affects the kinetic ordering of contact formation, for a protein with multiple viable folding pathways like CI2, HI will favor a particular folding route in a complex folding energy landscape. In that sense, ELT is short of fully predicting folding rates. From Secs. III B to III E, we investigate the cause of this temperature dependence of the effect of HI on folding rates and the implications for energy landscape theory. We also suggest a possible experimental design to probe the impact of HI on folding based on a temperature-dependent $\phi$-values analysis.

## II. MODELS AND METHODS

### A. Coarse-grained protein model

We used a coarse-grained, structure-based model [17] for two well-studied, model proteins: 64-residue $\alpha/\beta$ protein chymotrypsin inhibitor 2 (CI2) (PDB ID: 1YPA) [18] in Fig. 1(a), and the 57-residue $\beta$-barrel $\alpha$-spectrin Src-homology 3 (SH3) domain (PDB ID: 1SHG) [19] in Fig. 1(b). A structure-based model is a toy model that provide a single global basin of attraction that corresponds to an experimentally determined configuration and smooths out the ruggedness on the funneled energy landscape [26]. This allows us to study the ideal energy landscape of a protein. In this coarse-grained model, each residue is represented by one bead placed at its α-carbon position, creating a string of beads that represents the entire protein [see Figs. 1(c) and 1(d) for CI2 and SH3, respectively]. The Hamiltonian of our system depends on the experimental determined configuration (also known as the native state) consisting of backbone terms, attractive interactions between beads in close proximity to each other in the native state, and excluded volume, having the following form taken from the model developed by Clementi *et al.* [17]:

$$\mathcal{H}\left(\Gamma, \Gamma^0\right) = \sum_{i<j} k_r \left(r_{ij} - r_{ij}^0\right)^2 \delta_{j,i+1} + \sum_{i \in angles} k_\theta \left(\theta_i - \theta_i^0\right)^2$$
$$+ \sum_{i \in dihedral} k_\phi \left( \left\{ 1 - \cos\left[\phi_i - \phi_i^0\right] \right\} + \frac{1}{2} \left\{ 1 - \cos\left[3\left(\phi_i - \phi_i^0\right)\right] \right\} \right) \qquad (1)$$
$$+ \sum_{\substack{|i-j|>3 \\ i,j \in native}} \varepsilon \left[ 5 \left(\frac{r_{ij}^0}{r_{ij}}\right)^{12} - 6 \left(\frac{r_{ij}^0}{r_{ij}}\right)^{10} \right] + \sum_{\substack{|i-j|>3 \\ i,j \notin native}} \varepsilon \left(\frac{\sigma}{r_{ij}}\right)^{12},$$

where $\Gamma$ is a configuration of the set: $r$, $\theta$, $\phi$. The $r_{ij}$ term is the distance between $i$th and $j$th residues, $\theta$ is the angle defined by three consecutive beads, and $\phi$ is the dihedral angle between four consecutive beads. We define $\varepsilon = 0.6$ kcal/mol as the solvent-mediated interaction, and $k_r = 100\varepsilon$, $k_\theta = 20\varepsilon$, $k_\phi = \varepsilon$ [17]. $\delta$ is the Kronecker delta function. The native state values of $r^0, \theta^0$, and $\phi^0$ for both proteins were obtained from their crystal structures $\Gamma^0$ [18,19] where $\Gamma^0 = \{\{r^0\}, \{\theta^0\}, \{\phi^0\}\}$. The non-bonded terms consist of a Lennard-Jones interaction between native pairs, and excluded volume interaction between non-native pairs. The native contact pairs were chosen using the CSU program [27]. $\sigma$ for non-native pairs is 4 Å [17,28].

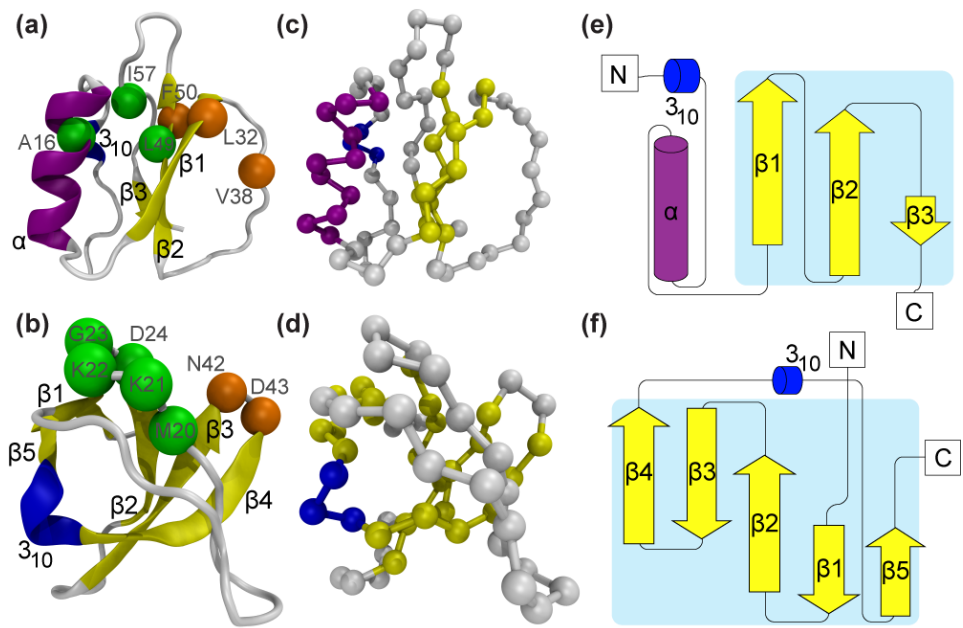

FIG. 1. Representations of protein models of Chymotrypsin inhibitor 2 (CI2) (top row) and the α-spectrin Src-homology 3 (SH3) domain (bottom row). The protein models are in [(a), (b)] a cartoon representation, [(c), (d)] a $C_{\alpha}$–only representation, and [(e), (f)] a protein topology cartoon created with Pro-origami [29]. Arrows are β-strands and cylinders are helices. Structures from [(a), (b), (c), (d)] were created with VMD [30]. The secondary structures were assigned using DSSP [31]. The key residues for the hydrophobic core (A16, L49, and I57) and the mini-core (L32, V38, and F50) are represented with green and orange beads, respectively, for CI2 in (a). Residues of the diverging turn (M20, K21, K22, G23, and D24) and distal loop (N42 and D43) are represented with green and orange for SH3 in (b).

A short description of both proteins is necessary for later results. CI2 has one α-helix packed against three β-strands (from β1 to β3), and a $3_{10}$-helix ($3_{10}$) as shown in Fig. 1(e). The two key cores are the hydrophobic core and the mini-core [the key residues are shown in Fig. 1(a)] [32,33]. SH3 has five β-strands (from β1 to β5), and a $3_{10}$-helix ($3_{10}$) as shown in Fig. 1(f). The β-strands are arranged antiparallel with respect to each other. Diverging turn and distal loop are key to the formation of the transition state configurations [see Fig. 1(b)] [34,35].

## B. Brownian dynamics with or without HI

Our protein folding simulations utilized a Brownian dynamics with HI (BDHI) method developed by Ermark and McCammon [25]. We used the software HIBD developed by the Skolnick group [36] using only the far-field hydrodynamic interactions. The equation of motion is given by

$$\mathbf{x}_i\left(t+dt\right) = \mathbf{x}_i\left(t\right) + \frac{\sum_j \mathbf{D}_{ij}\mathbf{F}_j}{k_{\mathrm{B}}T}dt + \mathbf{G}_i\left(dt\right), \tag{2}$$

where $\mathbf{x}_i\left(t+dt\right)$ is the position vector of the $i$th $C_\alpha$ bead at time $t+dt$. $\mathbf{F}_j$ is the total force acting on the $j$th $C_\alpha$ bead. The diffusion tensor $\mathbf{D}$ is a supermatrix of $3N{\times}3N$, where $N$ is the number of beads. $\mathbf{D}_{ij}$ is the $3{\times}3$ submatrix in the $i$th row and $j$th column of the diffusion tensor. In the absence of HI, the off-diagonal submatrices are zero. The diagonal terms were calculated from the Stokes-Einstein relation shown in Eq. (3) where $\eta$ is the viscosity of the aqueous solvent at temperature $T$ [36], $k_{\mathrm{B}}$ is the Boltzmann constant, $a$ represents the hydrodynamic radius of the beads (5.3 Å for each $C_\alpha$ residue is taken from [10]), and $\mathbf{I}_3$ is a $3{\times}3$ identity matrix. In the presence of HI, the elements of submatrices $\mathbf{D}_{ij}$ were obtained from the equations developed by Rotne and Prager [37] from the solutions of the Navier-Stokes equation under a low Reynolds number, and Yamakawa [38] extended the expression between a pair whose distance separation is less than the size of a bead. The complete set of formulas to compute $\mathbf{D}_{ij}$ terms is shown below in Eqs. (3-5). For an equation of motion with HI, then

$$\mathbf{D}_{ii} = \frac{k_{\mathrm{B}}T}{6\pi\eta a}\mathbf{I}_3, \tag{3}$$

$$\mathbf{D}_{ij} = \frac{k_{\mathrm{B}}T}{8\pi\eta r_{ij}}\left[\left(1+\frac{2}{3}\frac{a^2}{r_{ij}^2}\right)\mathbf{I}_3 + \left(1-\frac{2a^2}{r_{ij}^2}\right)\left(\frac{\mathbf{r}_{ij}\otimes\mathbf{r}_{ij}}{r_{ij}^2}\right)\right] \quad r_{ij}\ge 2a, \tag{4}$$

$$\mathbf{D}_{ij} = \frac{k_{\mathrm{B}}T}{6\pi\eta a}\left[\left(1-\frac{9}{32}\frac{r_{ij}}{a}\right)\mathbf{I}_3 + \frac{3}{32}\frac{\mathbf{r}_{ij}\otimes\mathbf{r}_{ij}}{ar_{ij}}\right] \quad r_{ij} < 2a, \tag{5}$$

where $\otimes$ represents a tensor product between two vectors. For the simulations using Brownian dynamics in the absence of HI (BD), the diffusion matrix reduces to Eqs. (6) and (7):

$$\mathbf{D}_{ii} = \frac{k_{\mathrm{B}}T}{6\pi\eta a}\mathbf{I}_3, \tag{6}$$

$$\mathbf{D}_{ij} = \mathbf{0}_3 \quad i\ne j. \tag{7}$$

$\mathbf{G}_i(dt)$ in Eq. (2) is the random displacement that mimics the stochastic behavior on a $C_\alpha$ bead from the implicit solvent. The relation between the random displacement and the diffusion tensor is linked by Eq. (8), which ensures that the fluctuation-dissipation theorem is satisfied. $\langle \, \rangle$ represents the ensemble average.

$$\langle \mathbf{G}_i(dt)\mathbf{G}_j(dt) \rangle = 6\mathbf{D}_{ij}dt \quad \text{and} \quad \langle \mathbf{G}_i(dt) \rangle = 0 \,. \tag{8}$$

### C. Equilibrium thermodynamic simulations

To evaluate the thermodynamic properties of CI2 and SH3, we utilized molecular dynamics simulation with BD and BDHI. To acquire sampling efficiency of the conformational space of the proteins, we used the replica exchange method (REM) [39] for enhanced sampling. The initial structures were chosen from an ensemble of unfolded structures that were annealed progressively until they reach the target temperature. For each protein, 20 temperatures were chosen for a set of REM simulation. The integration time step $dt$ is $10^{-3}\,\tau$, where $\tau = \sqrt{m\sigma_\alpha^2/\varepsilon}$, where $\sigma_\alpha$ is the average distance between two consecutive $C_\alpha$ beads (3.8 Å) and $m$ is 100 u representing the mass of a bead. The sampling rate was greater than the correlation time. The acceptance or rejection of each exchange between replicas follows the Metropolis criterion [40], $\min(1, \exp\{[\beta_i - \beta_j] \cdot [\mathcal{H}(\Gamma_i) - \mathcal{H}(\Gamma_j)]\})$, where $i$ and $j$ are two consecutives replicas, $\beta = 1/k_\mathrm{B}T$, $k_\mathrm{B}$ is the Boltzmann constant, $T$ is the temperature, and $\mathcal{H}$ is the potential energy of the system.

The number of samples was determined according to the convergence of the potential energy for all temperatures. Ensembles of $2.4\times10^5$ and $1.6\times10^5$ statistically significant conformations were obtained for each replica of CI2 and SH3, respectively. We computed thermodynamics properties and errors from the simulations with the weighted histogram analysis method (WHAM) [41].

### D. Non-equilibrium kinetic simulations

*1. Generation of the non-Arrhenius plot without HI*

We simulated folding kinetics using Brownian dynamics in the absence of HI (BD) for CI2 and SH3 over wide ranges of temperatures. We represent the temperature used for each protein in units of their corresponding folding temperature $T_f$, where they are defined as the temperatures when the free energy (FE), with respect to the fraction of native contact formation $Q$, of the unfolded state is equal to the FE of the folded state; i.e., $FE(Q_u) - FE(Q_f) = 0$ where the basin of the unfolded state $Q_u$ and folded states $Q_f$ are equal in the free energy. Thus, temperatures are represented in units of $T_f^{CI2}$ and $T_f^{SH3}$ for CI2 and SH3, respectively. We explored the range of temperatures from $0.1\,T_f^{CI2}$ to $1.3\,T_f^{CI2}$ for CI2 and from $0.1\,T_f^{SH3}$ to $1.25\,T_f^{SH3}$ for SH3. A folding kinetic simulation started from an unfolded configuration, which was chosen randomly from an ensemble at high temperature, was performed until it reached the folded state for the first time (first passage time). We considered that a protein is folded when the non-bonded potential, the sum of the last two terms of Eq. (1), is less or equal to 0.9 of the native state non-bonded potential for CI2 and SH3. The number of trajectories depends on the convergence of the mean first passage time (MFPT) at each temperature. The average folded time $t_{fold}$ is the MFPT. The maximum simulation time $t_{max}$ is $9 \times 10^6\,\tau$ and was chosen as the folding time for trajectories that did not reach the folded state.

### *2. The impact of HI on protein folding kinetics*

We selected another range of temperatures for simulations with HI for each protein, from $0.95\,T_f^{CI2}$ to $1.06\,T_f^{CI2}$ for CI2 and from $0.91\,T_f^{SH3}$ to $1.03\,T_f^{SH3}$ for SH3. The number of trajectories from unfolded to the folded state depends on the convergence of MFPT at each condition. The maximum simulation time $t_{max}$ is $9 \times 10^6\,\tau$ and was chosen as the folding time for trajectories that did not reach the folded state. For the analysis where the kinetics trajectories were projected on a two dimensional energy landscape (see Sec. III E), the number of trajectories were increased to 1500 and 2500 to reduce statistical error for CI2 and SH3 trajectories, respectively.

### **E. Effective diffusion coefficient of a reaction coordinate**

We expect changes in the diffusion coefficient of a reaction coordinate to reflect the changes in the predictions of folding rates from the energy landscape theory (ELT) because HI is a kinetic effect that will not alter the overall free energy profiles [i.e., Hamiltonian is the same with or without HI, see Eq. (1)]. Thus, instead of comparing the analytically predicted folding rates, we computed an effective diffusion coefficient along a reaction coordinate, the fraction of the native contact formation $Q$. We used the following expression to fit $D^{eff}$ from mean square displacement (MSD) of $Q$ over a lag time $t'$ [42]:

$$D^{eff} = \frac{1}{2d} \lim_{t' \to t_D} \frac{\left\langle \left[ Q(t_0 + t') - Q(t_0) \right]^2 \right\rangle_{t_0, \Omega}}{t'},$$  (9)

where $\langle \rangle_{t_0, \Omega}$ represents the average over all simulated time $t_0$ separated by a lag time $t'$ and all trajectories $\Omega$. $d$ is the dimension of one. $t_D$ is the timescale where the MSD shows a linear behavior with time. We followed Whitford's method to compute $D^{eff}$ [42] by fitting the diffusion coefficient at the regime where MSD varies linearly with lag time.

The ensemble average of the MSD is performed over a selected number of kinetic trajectories. The MSD is measured from the unfolded basin ($Q_u$ =0.2 and $Q_u$ =0.1 for CI2 and SH3, respectively) to slightly above the top of the barrier ($Q^{\ddagger}$ =0.6 and $Q^{\ddagger}$ =0.5, for CI2 and SH3, respectively).

### F. Data analysis

#### 1. Differences in the probability of secondary structure formation

The probability of secondary structure formation as a function of time $P(t)$ taken over all kinetic trajectories is estimated in the presence or in the absence of HI for both proteins. The differences in the probability of secondary structure formation between BDHI and BD ($\Delta P(t)$) is defined as:

$$\Delta P(t) = P^{BDHI}(t) - P^{BD}(t).$$  (10)

When $\Delta P(t)$ is positive, the average probability of secondary structure formation from BDHI is greater than BD, and vice versa when $\Delta P(t)$ is negative.

### 2. Displacement correlation

We calculated the displacement vector of a residue $k$, which is defined as $\mathbf{s}_k(t) = \mathbf{x}_k(t) - \mathbf{x}_k(t - dt)$, throughout the kinetic simulation. Then, we investigated the displacement correlation $C_{ij}(t)$ by taking the cosine of the angle formed between two displacement unit vectors $\hat{\mathbf{s}}(t)$ of a pair of residues $i$ and $j$ ($i{\neq}j$) at time $t$ and averaging over all trajectories $\Omega$ as:

$$C_{ij}(t) = \left\langle \hat{\mathbf{s}}_i(t) \cdot \hat{\mathbf{s}}_j(t) \right\rangle_{\Omega}. \tag{11}$$

The translation and rotation of the center of mass of each configuration was removed before calculating the displacement correlation.

### 3. Chance of occurrence

The chance of occurrence $CoO(|i-j|)$ is defined as the ratio of number of residue pairs with a sequence separation $|i\text{-}j|>0$ whose magnitude of the displacement correlation is above a selected threshold, and the total number of residue pairs at that sequence separation, as:

$$CoO(|i-j|) = \frac{\sum\limits_{i'>j'} \Theta(C_{i'j'} - \mu)\, \delta_{|i'-j'|,|i-j|}}{\sum\limits_{i'>j'} \delta_{|i'-j'|,|i-j|}}, \tag{12}$$

where $\Theta$ is the Heaviside step function and $\delta$ is the Kronecker delta function. The chosen threshold $\mu$ is the average positive displacement correlation from the ensemble.

$$\mu = \frac{\sum\limits_{i>j} C_{ij}\, \Theta(C_{ij})}{\sum\limits_{i>j} \Theta(C_{ij})}. \tag{13}$$

Negative displacement correlation is ignored because the signal is not as strong.

### III. RESULTS

## A. The impact of BDHI on the folding time depends on temperature

We explored the folding kinetics of CI2 and SH3 by comparing the mean first passage time (MFPT) with BD over a broad range of temperatures. Both proteins exhibit non-Arrhenius [43] behavior against temperature as shown in Figs. 2(a) and 2(b). At high temperatures, the MFPT increases because the thermal fluctuations are higher than the stability of the protein, and at low temperatures, the MFPT increases due to the fact that the protein is trapped in a local energy minimum [15,43,44]. The temperature that renders the fastest MFPT is at $0.95\,T_f^{CI2}$ and $0.91\,T_f^{SH3}$ for CI2 and SH3, respectively. We computed the MFPT for the proteins with BDHI over a narrow range of temperatures around $T_f$ of the proteins in Figs. 2(c) and 2(d) in dashed lines. Our study shows that the impact of HI on the MFPT is small within an order of magnitude, but statistically significant. What is most interesting is that HI either increases or decreases the folding time depending whether the temperature is higher or lower than $T_f$. This distinctive "crossover" behavior occurs in the proximity of the folding temperature of CI2 ($\approx 1.03\,T_f^{CI2}$) and SH3 ($\approx 0.98\,T_f^{SH3}$). Thus, the impact of HI on protein folding kinetics is temperature dependent. However, the acceleration of the folding is more prominent for CI2 than for SH3 at $T < T_f$. Therefore, HI effects also depend on the topology of a protein. We will further investigate the role of topology in the extent of impact from HI on protein folding in the following subsection at two temperatures for each protein: below $T_f$ ($0.95\,T_f^{CI2}$ for CI2 and $0.91\,T_f^{SH3}$ for SH3) and above $T_f$ ($1.06\,T_f^{CI2}$ for CI2 and $1.03\,T_f^{SH3}$ for SH3).

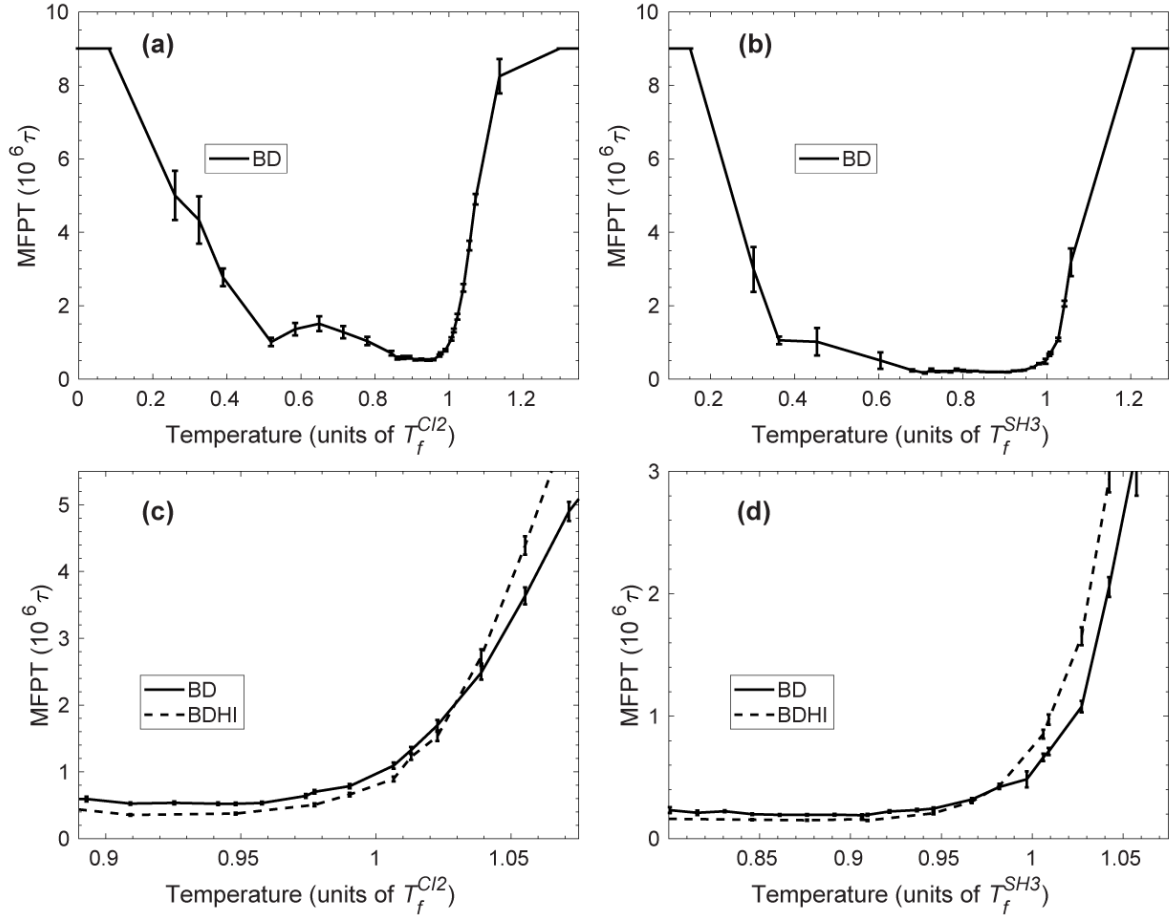

FIG. 2. The mean first passage time (MFPT) with respect to temperature for CI2 and SH3 in the presence or in the absence of hydrodynamic interactions (HI). Panels (a) and (b) show the MFPT over a broad range of temperatures using the Brownian dynamics without HI (BD) for CI2 and SH3, respectively. The temperature for each protein is expressed in units of their corresponding folding temperature $T_f$. Note the U-shaped dependence of the folding time (non-Arrhenius behavior). The MFPT using BD with HI (BDHI) is compared to the MFPT using only BD in panel (c) for CI2 and (d) for SH3. The crossover occurs when the two curves intersect. Error bars are calculated using the jackknife method.

## B. The effective diffusion coefficient of $Q$ partly accounts for the crossover behavior of the folding kinetics

Can we capture this folding behavior using a global order parameter? A theoretical estimation of the folding kinetic rate $k$ (the rate is the inverse of folding time $t_{fold}$) depends on the shape of free energy surface and the effective diffusion coefficient $D^{eff}$ of an order parameter on the free energy surface [43-45] as such,

$$k = \frac{1}{t_{fold}} = \left(\frac{\beta}{2\pi}\right)^{1/2} D^{eff} \omega \omega^{\ddagger} \exp(-\beta \Delta F^{\ddagger}), \qquad (14)$$

where $\omega$ and $\omega^{\ddagger}$ are the curvatures of the unfolded state free energy well and barrier, respectively, $\beta$ is the inverse temperature, and $\Delta F^{\ddagger}$ is the free energy barrier height with respect to the unfolded state free energy. However, since the Hamiltonian for BD and BDHI are identical rendering the same free energy profiles (see Fig. 3), the change in the folding kinetic rates should be explained by the change in the diffusion of the order parameter. Here, the order parameter is the fraction of native contact formation $Q$. The mean square displacement (MSD) of $Q$ is obtained as a function of time as shown in Fig. 4 to estimate the effective diffusion coefficients. The trajectories longer than $2\times10^5 \, \tau$ at $0.95 T_f^{CI2}$ and $8\times10^5 \, \tau$ at $1.06 \, T_f^{CI2}$ are considered for CI2, and $1.6\times10^5 \, \tau$ at $0.91 T_f^{SH3}$ and $4\times10^5 \, \tau$ at $1.03 T_f^{SH3}$ for SH3. The MSD is calculated for a time shorter than the average folding time for each condition. The initial phase is a transitional sub-diffusive process characterized by MSD~$t^{\alpha}$ with $\alpha<1$. After the memory from the initial state dissipates [46,47], the MSD reaches a normal diffusive regime. $D^{eff}$ of $Q$ is estimated from the linear region of the MSD of $Q$ as a function of lag time as shown in the insets of Fig. 4. A smaller diffusion coefficient at $T>T_f$ than that of $T<T_f$ is simply a thermal argument.

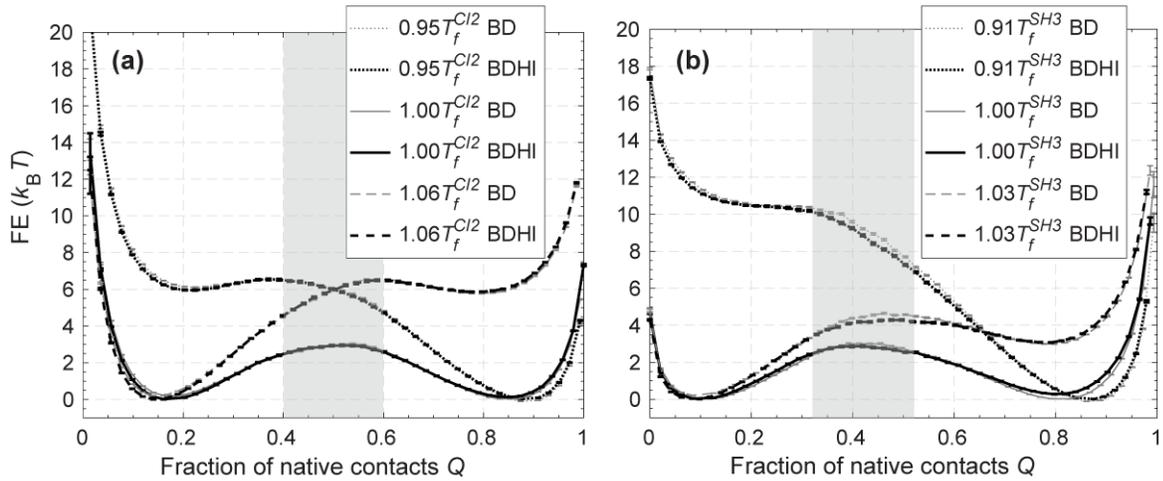

FIG. 3. Free energy (FE) with respect to fraction of native contact formation $Q$ without or with HI (BD and BDHI, respectively) for (a) CI2 and (b) SH3 at a temperature below the folding temperature ($T_f$), at $T_f$, and above $T_f$. Error bars are included.

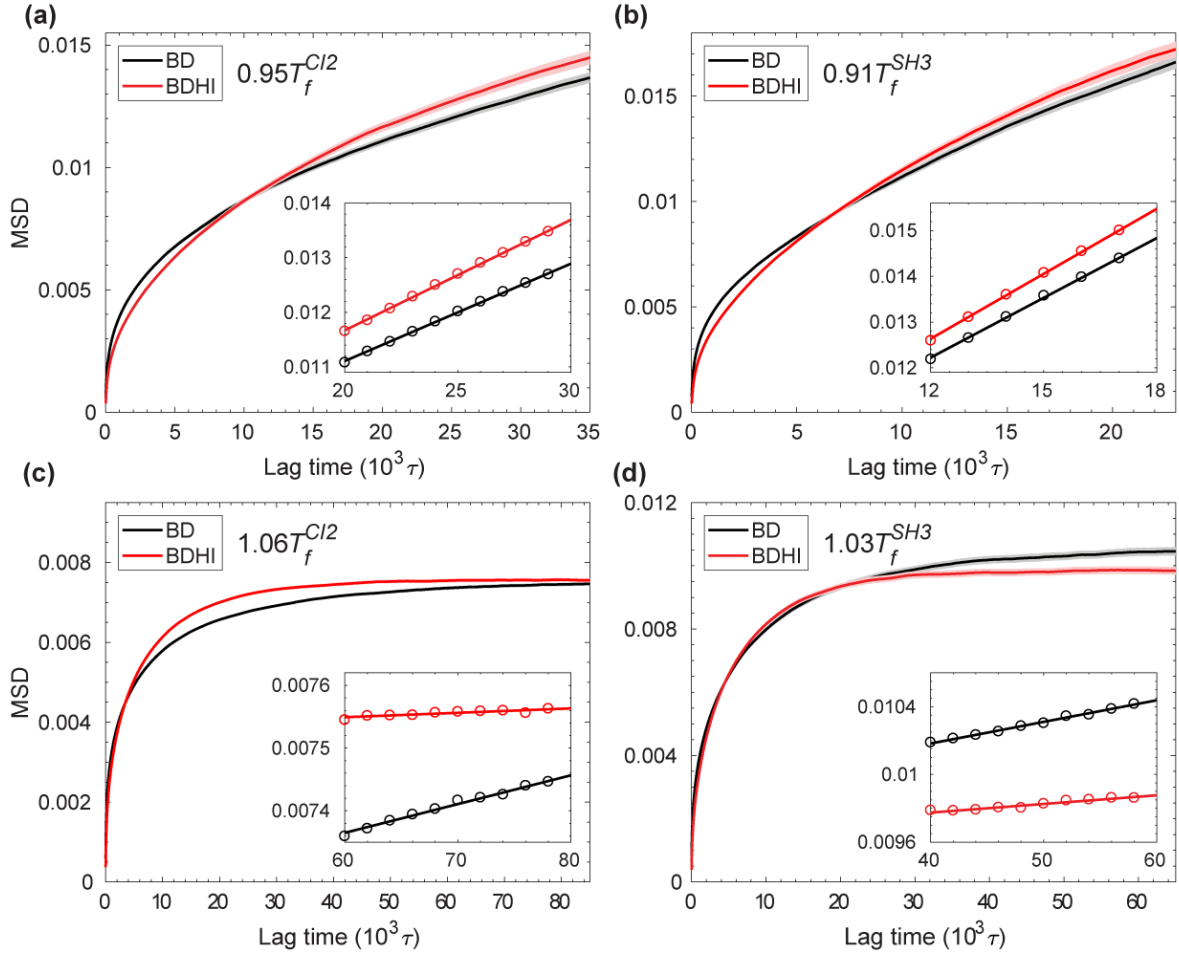

FIG. 4. The mean square displacement (MSD) of the fraction of the native contact formation $Q$ as a function of lag time in the absence or presence of HI (BD and BDHI, respectively) at $T<T_f$ [(a) $0.95T_f^{CI2}$ and (b) $0.91T_f^{SH3}$ for CI2 and SH3, respectively], and $T>T_f$ [(c) $1.06T_f^{CI2}$ and (d) $1.03T_f^{SH3}$ for CI2 and SH3, respectively]. Shaded width of solid lines represent the error. The inset zooms in to the range of time used for a linear fit (solid line) of data at every 100 ($T<T_f$) or 200 ($T>T_f$) points in open circles. The unit for the effective diffusion coefficient is $1/\tau$ since $Q$ is dimensionless.

If $Q$ is a good reaction coordinate that captures the dynamics over the entire free energy landscape, the ratio of the folding rates $k_{BDHI}/k_{BD}$ computed from kinetic simulations should be equal with the same ratio computed from MSD analysis. A comparison of the ratios of the folding rates $k_{BDHI}/k_{BD}$ computed from kinetic simulations with the same ratio computed from MSD analysis are shown in Table I and II for CI2 and SH3, respectively. The ratio $D_{BDHI}^{eff}/D_{BD}^{eff}$, however, is not equal to the ratio $t_{fold}^{BD}/t_{fold}^{BDHI}$ ; although it shows the right trend of the crossover behavior. For CI2, the ratio of folding rates from BDHI and BD kinetic simulations is $k_{BDHI}/k_{BD} = 1.37$ at $0.95T_f^{CI2}$ and $k_{BDHI}/k_{BD} = 0.83$ at $1.06T_f^{CI2}$, whereas the ratio of $k_{BDHI}/k_{BD}$ from fitting the MSD is $1.14$ at $0.95T_f^{CI2}$ and $0.15$ at $1.06T_f^{CI2}$. A similar trend of crossover behavior is observed for SH3. $D_{BDHI}^{eff}/D_{BD}^{eff}$ for either CI2 or SH3 is merely around 1 at $T<T_f$, while the $t_{fold}^{BD}/t_{fold}^{BDHI}$ computed from the folding simulations is above 1.18. The analysis of the diffusivity shows a retarded dynamics due to HI at $T>T_f$. The results from kinetic simulation also show retarded dynamics but less so. We speculate that a mean-field description of overall folding with the collective order parameter of $Q$ along an energy landscape may not fully grasp the kinetic principle of HI on folding. We will investigate the influence of HI on folding by the formation of the local contacts or secondary structures in the next subsection.

TABLE I. Folding time from kinetic simulations ($t_{fold}$) and the effective diffusion coefficient ($D^{eff}$) of $Q$ for CI2 using BD or BDHI.

| $T$ (in units of $T_f^{CI2}$) | From kinetic simulations | | | From MSD analysis | | |
|---|---|---|---|---|---|---|
| | $t_{fold}^{BD}$ ($10^6\,\tau$) | $t_{fold}^{BDHI}$ ($10^6\,\tau$) | $k_{BDHI}/k_{BD} =$ $t_{fold}^{BD}/t_{fold}^{BDHI}$ | $D_{BD}^{eff}$ ($10^{-9}\,1/\tau$) | $D_{BDHI}^{eff}$ ($10^{-9}\,1/\tau$) | $k_{BDHI}/k_{BD} =$ $D_{BDHI}^{eff}/D_{BD}^{eff}$ |
| 0.95 | $0.52 \pm 0.02$ | $0.38 \pm 0.01$ | $1.37 \pm 0.06$ | $89.16 \pm 0.08$ | $101.22 \pm 0.09$ | $1.14 \pm 0.00$ |
| 1.06 | $3.64 \pm 0.13$ | $4.39 \pm 0.14$ | $0.83 \pm 0.04$ | $2.29 \pm 0.01$ | $0.35 \pm 0.00$ | $0.15 \pm 0.00$ |

TABLE II. Folding time from kinetic simulations ($t_{fold}$) and the effective diffusion coefficient ($D^{eff}$) of $Q$ for SH3 using BD or BDHI.

| $T$ (in units of $T_f^{SH3}$) | From kinetic simulations | | | From MSD analysis | | |
|---|---|---|---|---|---|---|
| | $t_{fold}^{BD}$ ($10^6\,\tau$) | $t_{fold}^{BDHI}$ ($10^6\,\tau$) | $k_{BDHI}/k_{BD} =$ $t_{fold}^{BD}/t_{fold}^{BDHI}$ | $D_{BD}^{eff}$ ($10^{-9}\,1/\tau$) | $D_{BDHI}^{eff}$ ($10^{-9}\,1/\tau$) | $k_{BDHI}/k_{BD} =$ $D_{BDHI}^{eff}/D_{BD}^{eff}$ |
| 0.91 | $0.19 \pm 0.01$ | $0.16 \pm 0.01$ | $1.19 \pm 0.10$ | $217.72 \pm 0.30$ | $236.71 \pm 0.24$ | $1.09 \pm 0.00$ |
| 1.03 | $1.08 \pm 0.05$ | $1.65 \pm 0.07$ | $0.65 \pm 0.04$ | $6.39 \pm 0.01$ | $2.56 \pm 0.01$ | $0.40 \pm 0.00$ |

## C. HI facilitates the ordering of key structural regions at $T<T_f$

In the previous subsection, we have shown how HI impacts folding globally to explain the crossover behavior of the folding rates; however, HI also impacts local secondary structure formation. We investigated the temperature dependence of HI and the crossover behavior by analyzing the ordering of secondary structures of the proteins along a time that is normalized by the maximum time $t_{max}$. For the selected temperatures, we calculated the differences in the probability of secondary structure formation $\Delta P(t)$ between BDHI and BD as a function of normalized time (Fig. 5). For each protein, one temperature is slightly below $T_f$ [Figs. 5(a) and 5(b) for CI2 and SH3, respectively] and the other is slightly above $T_f$ [Figs. 5(c) and 5(d) for CI2 and SH3, respectively]. At $T<T_f$ the folding time with BDHI decreases with respect to BD, and vice versa at $T>T_f$. We are interested at analyzing the folding formation at a time before the proteins reach the transition state at the top of the folding barrier. The transition state region is shown in the grey shaded area for each protein in Fig. 3, and it is a key part in the folding process. This stage of folding occurs before the dashed, vertical lines in each panel of Fig. 5 (see Fig. 6 for a complete temporal evolution of $\langle Q \rangle_\Omega$ along normalized time where $\langle \ \rangle_\Omega$ represents the average over all kinetic trajectories).

The two temperatures to investigate CI2's crossover behavior are $0.95\,T_f^{CI2}$ and $1.06\,T_f^{CI2}$. We grouped the native contact pairs into secondary structure segments to get a structural view of the impact of HI. Figure 5(a) shows that at $T<T_f$, the most positive $\Delta P(t)$ is observed for β1-β2 and β1-seg4 that form the mini-core, and seg3-β2, seg3-seg4, seg4-β2 that are in the neighborhood of the mini-core. It shows a modest positive $\Delta P(t)$ for $3_{10}$-α (native pairs close to the N-terminus) and β2-β3 (native pairs close to the C-terminus). This suggests that BDHI enhances the formation of the secondary structures within the mini-core and have less impact at the termini. In Fig. 5(c) at $T>T_f$, the native contacts that are more affected by BDHI than by BD [negative $\Delta P(t)$] are the segments involving seg3-seg4, seg4-β2, the long-range contacts $3_{10}$-seg5, and contacts in the C-terminus β2-β3. This implies that the impact of HI has a longer range at $T>T_f$ than at $T<T_f$.

Turning the attention to SH3, the two temperatures to investigate SH3's crossover behavior are $0.91T_f^{SH3}$ and $1.03T_f^{SH3}$. At $T<T_f$ [Fig. 5(b)], the most positive $\Delta P(t)$ is observed in the region about the RT loop (RT), the diverging turn (DT), and the distal loop (DL). Those involved secondary structures are DT-$\beta$4, DT-$\beta$5, DT-$3_{10}$, RT-$\beta$3, RT-$\beta$4, and N-src-$\beta$5. BDHI promotes specific contacts in SH3 that is known for its obligatory role in the formation of the transition state. In addition, the contacts in the neighborhood of the distal loop and N-src loop are also mildly enhanced with BDHI ($\beta$2-$\beta$3, $\beta$2-$\beta$4, $\beta$3-$\beta$5, and DT-$\beta$3). On the other hand, at $T>T_f$ [Fig. 5(d)] the formation of the previous pairs mentioned before are also negatively affected by BDHI indicated by a negative $\Delta P(t)$. However, the impact of HI on the folding time of SH3 is less than that of CI2. We will discuss the difference in the impact from the viewpoint of protein topology in the following subsections.

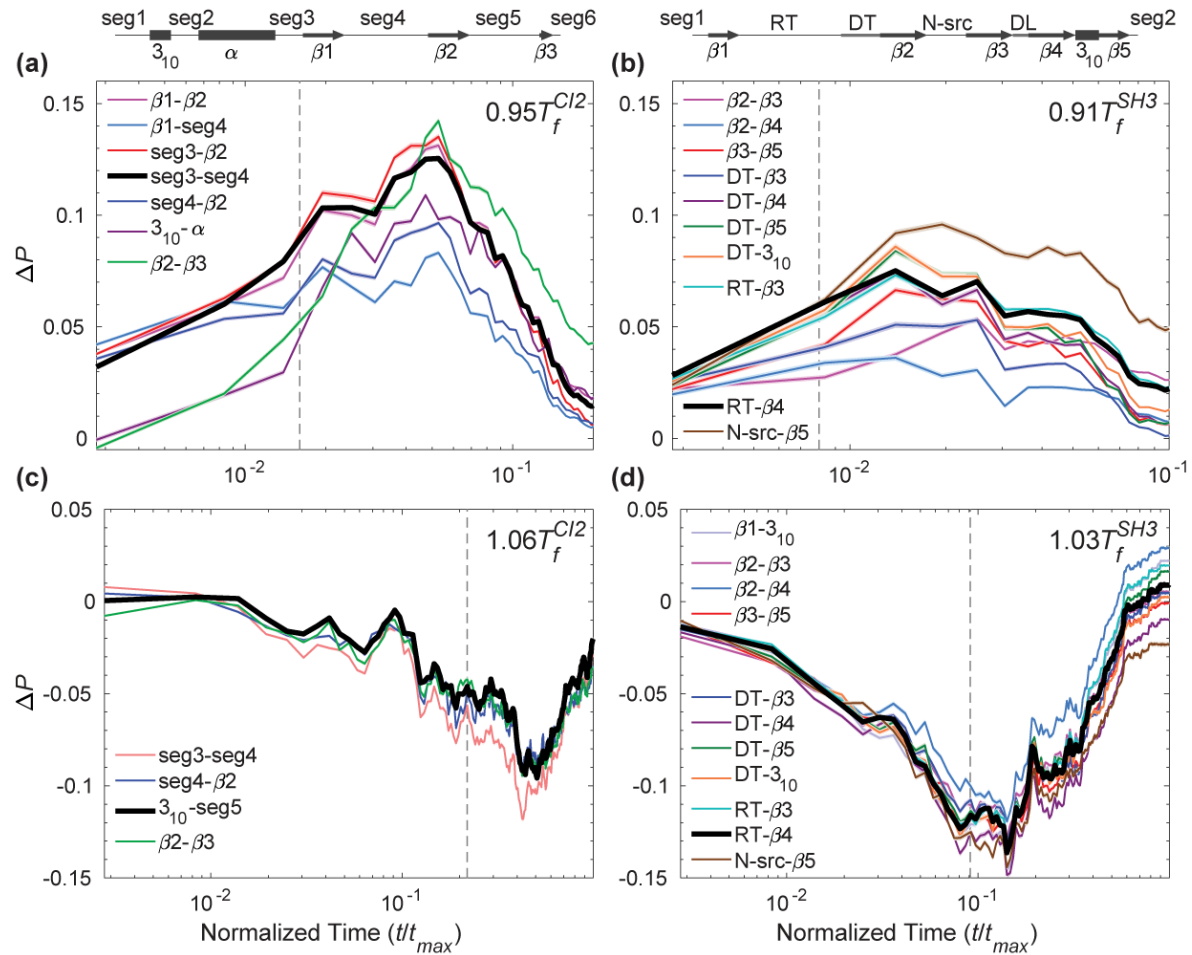

FIG. 5. Impact of HI on key pairs of secondary structures interactions $\Delta P(t)$ at $T < T_f$ [(a) $0.95\,T_f^{CI2}$ and (b) $0.91\,T_f^{SH3}$ for CI2 and SH3, respectively], and $T > T_f$ [(c) $1.06\,T_f^{CI2}$ and (d) $1.03\,T_f^{SH3}$ for CI2 and SH3, respectively]. Time is normalized with respect to maximum simulation time ($9 \times 10^6\,\tau$) and shown in log scale. The dashed, vertical lines correspond to the time when $\langle Q \rangle_\Omega = 0.4$ for CI2 and SH3 which is the $Q$ value in the transition state. The curves were smoothed by averaging every 500 data points. Errors are included but too small to be visible. At the top of the panels, the location of secondary structure elements and unstructured regions along the sequence of CI2 (left) and SH3 (right) is indicated as visual guidance.

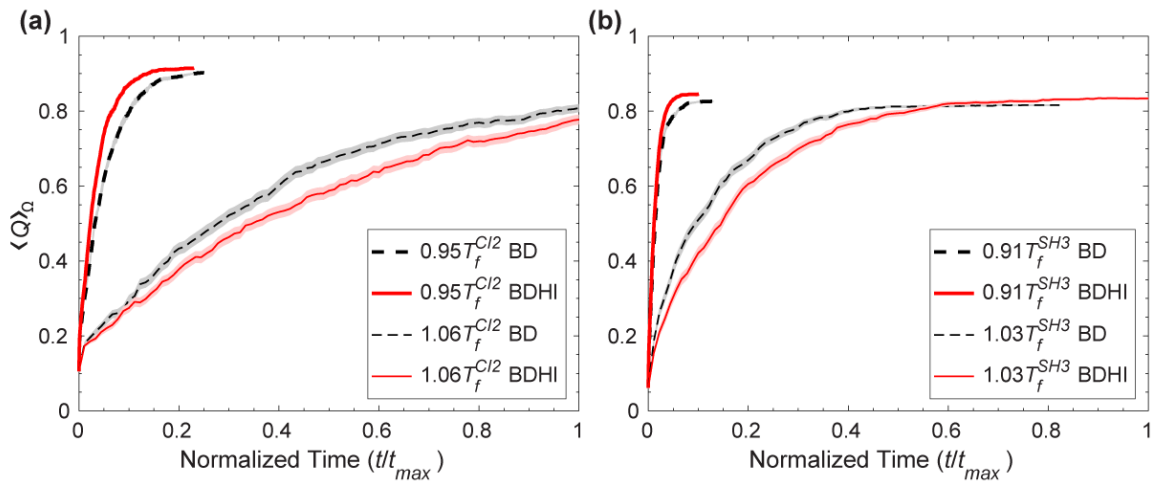

FIG. 6. Evolution of the average of the fraction of native contact formation $Q$ over all trajectories as function of normalized time $t/t_{max}$ at $T < T_f$ and $T > T_f$ for (a) CI2 and (b) SH3. Time is normalized with respect to maximum simulation time ($9 \times 10^6\,\tau$). Shaded width of lines represent the error, which is calculated using the jackknife method.

## D. Hydrodynamic coupling of mid-range and long-range contacts and their opposing impact on the general ordering of contact formation at $T < T_f$ and $T > T_f$

Armed with the knowledge of the kinetic specific structural formation from the previous subsection, we hypothesize that HI influences the self-organization of a protein configuration during the course of folding from an unfolded state to the formation of a transition state. Consequently, HI creates an opposing effect on folding time at a watershed of $T_f$, which is the crossover behavior in the folding kinetics. The addition of HI to the equation of motion introduces the many-body coupling between all beads where their motions are correlated inversely with their spatial separation $r$. To test this hypothesis, we compare the configurations in terms of probability of contact formation of each native pair $Q_{ij}(t)$ to the displacement correlation between a pair of contacts $C_{ij}(t)$ at a particular moment in folding kinetics. We chose this moment to be a time where the average of probability of contact formation $\langle Q \rangle_\Omega$ is 0.4 (see Fig. 6 for a temporal evolution of $\langle Q \rangle_\Omega$) as shown in Figs. 7 and 8 for CI2 and SH3, respectively .

The values of the displacement correlation $C_{ij}(t)$ at a particular time are shown in the lower triangles in Figs. 7 and 8 for CI2 and for SH3, respectively. The comparison of $C_{ij}(t)$ with $Q_{ij}(t)$ (upper triangles in Figs. 7 and 8) establishes a causal relationship between the dispersity of $Q_{ij}(t)$ and the spatial pattern of $C_{ij}(t)$. At $T < T_f$, the probability of contact formation among mid- to long-range pairs ($|i\text{-}j| \geq 10$) is disperse [upper triangle in Figs. 7(a) and 8(a)]. Their Fano factors (variance over mean) are 0.04 for CI2 [Fig. 7(b)] and 0.06 for SH3 [Fig. 8(b)]. Whereas at $T > T_f$, the probability of contact formation for pairs with $|i\text{-}j| \geq 10$ has a narrower distribution [upper triangle in Figs. 7(c) and 8(c)] evident by a lower Fano factor: 0.01 for CI2 [Fig. 7(d) and 0.02 for SH3 [Fig. 8(d)]. A rather high Fano factor implies the existence of certain localized contact formation when a protein folds from an unfolded state at $T < T_f$. A narrow dispersion [Figs. 7(d) and 8(d)] shows the contact formation is quite random as the protein folds and unfolds at $T > T_f$.

*1. CI2*

For CI2, the localized contacts are around the mini-core (highlighted in purple boxes). As shown in the lower triangle of Figs. 7(a) and 7(c) (at $T < T_f$ and $T > T_f$, respectively), HI alters the pattern of motions for long-range contacts (black boxes) and thus impacts the dynamics of mid-range contacts (green boxes). At $T > T_f$, the paired residues move cooperatively in the same direction; thus, adversely affecting the formation of the mid-range contacts around the mini-core.

Noticeably, in addition to the native pairs, the surrounding non-native pairs nearby are also correlated, which is not observed in the simulations with BD (Fig. 9). Several research groups have shown the importance of non-native pairs dictating protein kinetics [48-50]. To better visualize the whereabouts of the mid- to long-range contact pairs involving both native and non-native pairs, we projected the pairs with sequence separation of $|i\text{-}j| > 8$ on the native structure of CI2 in Fig. 7(e) with colored edges. The contact pairs with a similar range of positive correlation at both temperatures ($0.95\,T_f^{CI2}$ and $1.06\,T_f^{CI2}$) are shown with green edges. Most of which are located in the two regions formed by β1 β2 and β3, and the C-terminus with the connecting loop of β2 and β3. The pairs in which the magnitude of the correlation is greater at $0.95\,T_f^{CI2}$ than that of $1.06\,T_f^{CI2}$ are shown with blue edges, which are located mostly between the α-helix and the N-terminus, and between the connecting loop of β2 and β3 and the C-terminus. The pairs in which magnitude of the correlation is greater at $1.06\,T_f^{CI2}$ than that of $0.95\,T_f^{CI2}$ are shown with red edges. They are present between the N-terminus and the C-terminus, the α-helix and the following loop, and the region of the mini-core.

As hinted in the previous paragraph, we speculate the sequence separations between contacts, involving both native and non-native contacts, play a significant role in the crossover behavior in the presence of HI. To extend our analysis and justify our speculation, we plotted the chance of occurrence $CoO(|\,i-j\,|)$ (see definition in Sec. II F) in Fig. 7(f) along the sequence separation |$i$-$j$|. There is a stronger signal at mid-range contacts ($10<|i\text{-}j|<30$) at $T<T_f$ than that of $T>T_f$. Most noticeably, there is a strong signal at $|i\text{-}j|\approx 60$ that shows that long-range contacts are indeed correlated at $T>T_f$.

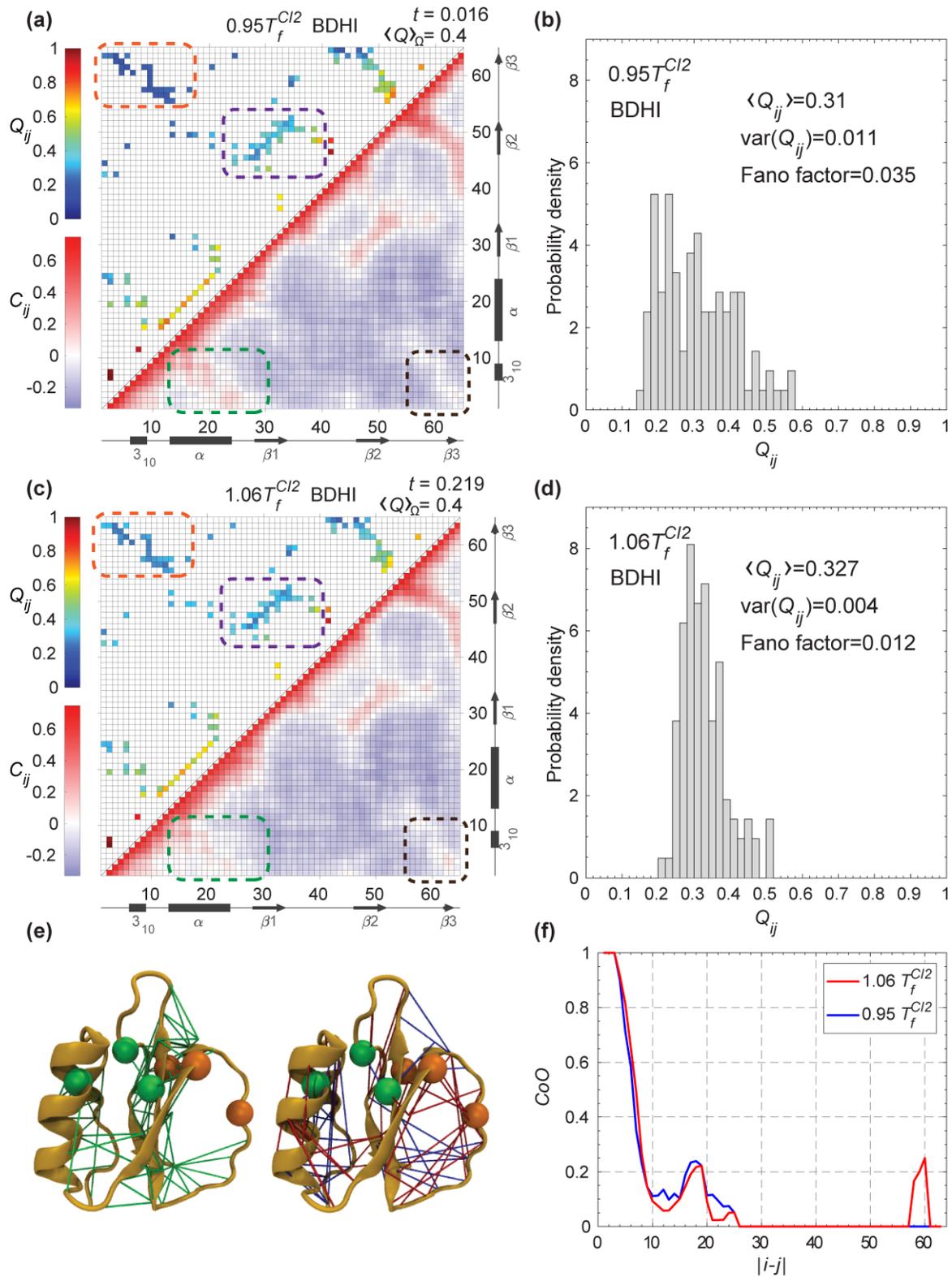

FIG. 7. (a) The probability of contact formation of each native pair $Q_{ij}(t)$ (upper triangle) and the displacement correlation $C_{ij}(t)$ (lower triangle) at $0.95T_f^{CI2}$ for CI2. (b) The distribution of native pairs from panel (a) with $|i\text{-}j|{\geq}10$. (c) $Q_{ij}\left(t\right)$ and $C_{ij}\left(t\right)$ at $1.06T_f^{CI2}$. (d) The distribution of native pairs from panel (c) with $|i\text{-}j|{\geq}10$. (a) and (c) correspond to a normalized time where $\langle Q\rangle_\Omega$ =0.4. The arrows and rectangles along (a) and (c) represent β-strands and helices, respectively. The orange and purple dashed boxes highlight the long-range contacts between the N-terminus and C-terminus, and the contacts in neighborhood of the mini-core, respectively. The black and green dashed boxes highlight the displacement correlation between the N-terminus and C-terminus, and between the N-terminus and the α-helix, respectively. (e) The displacement correlation at $0.95T_f^{CI2}$ and $1.06T_f^{CI2}$ for all pairs are classified into three sets based on the magnitude of positive correlations. If the magnitude of the correlation is similar at both temperatures, the pair is colored with a green edge on the left structure. If the magnitude of the correlation is greater at $0.95T_f^{CI2}$ than that at $1.06T_f^{CI2}$ a pair is colored with a blue edge on the right structure. If the magnitude of the correlation is greater at $1.06T_f^{CI2}$ than that at $0.95T_f^{CI2}$ a pair is colored with a red edge on the right structure. Only the pairs with sequence separation greater than 8 residues and magnitude of displacement correlation above the threshold $\mu$ =0.061 are considered for this representation. The key residues for the hydrophobic core (A16, L49, and I57) and the mini-core (L32, V38, and F50) are illustrated with green and orange beads, respectively. (f) $CoO(|i-j|)$ for all pairs whose magnitude of the displacement correlation is above $\mu$ are organized according to the sequence separation $|i\text{-}j|$.

### *2. SH3*

We found that HI affects SH3 (Fig. 8) in a similar way to CI2; however, the effect is not as strong. This is evident by the data collected at the time that corresponds to the transition state ($\langle Q\rangle_\Omega$ =0.4). The contact formation is localized at mid-range contacts between the diverging turn (DT) and the distal loop (DL) (in purple boxes), which is known to be critical to the formation of transition state ensemble experimentally [34,35]. Similar to Fig. 7(e), we projected the pairs with displacement correlations greater than the average positive correlation on the native structure in Fig. 8(e). Any pairs with sequence separation of $|i\text{-}j|$>7 are grouped in colored edges. The green ones correspond to the similar magnitude of pair correlation at a temperature either higher or lower than $T_f$. The pairs in which the magnitude of the correlation is greater at $0.91T_f^{SH3}$ than that of $1.03T_f^{SH3}$ are shown with blue edges, which are the pairs in the region of the RT and DT loop, 3$_{10}$-helix and β3. Furthermore, the pairs in which the magnitude of the correlation is greater at $1.03T_f^{SH3}$ than that of $0.91T_f^{SH3}$ are shown with red edges, which are the pairs between β2 and the N-terminus (from seg1 to DT), and the long-range pairs between seg1 and seg2. Again, we plotted the $CoO(|i-j|)$ in Fig. 8(f) along sequence separation for SH3. Similar to CI2, there is a stronger signal at mid-range contacts ($10<|i\text{-}j|<20$) at $T<T_f$ than that of $T>T_f$. At $|i\text{-}j|\approx56$, there are long-range contacts that are correlated at $1.03\,T_f^{SH3}$.

The previous analysis is compared to the same plots without HI (BD) in Fig. 9. Although the contact maps are similar to their corresponding $\langle Q \rangle_\Omega$ for both proteins, there are no clear pattern in the displacement correlation map for BD. The displacement correlation randomly fluctuates around zero. The $CoO(|i-j|)$ showed in Figs. 7(f) and 8(f) suggest that the crossover behavior in the presence of HI is due to the displacement correlation between the mid-range contacts and long-range contacts. The mid-range contacts for CI2 are the ones that form the mini-core, and for SH3 are the ones between the diverging turn and distal loop. Although we employed a structure-based model where the native pairs are energetically attractive, we identifed the importance of dynamic correlation between residues that form a native pair and their neighboring non-native pairs particularly between the α-helix and the N-terminus for CI2, and between both DT and β2 and N-terminus for SH3, for the retardation of the folding time at $T>T_f$.

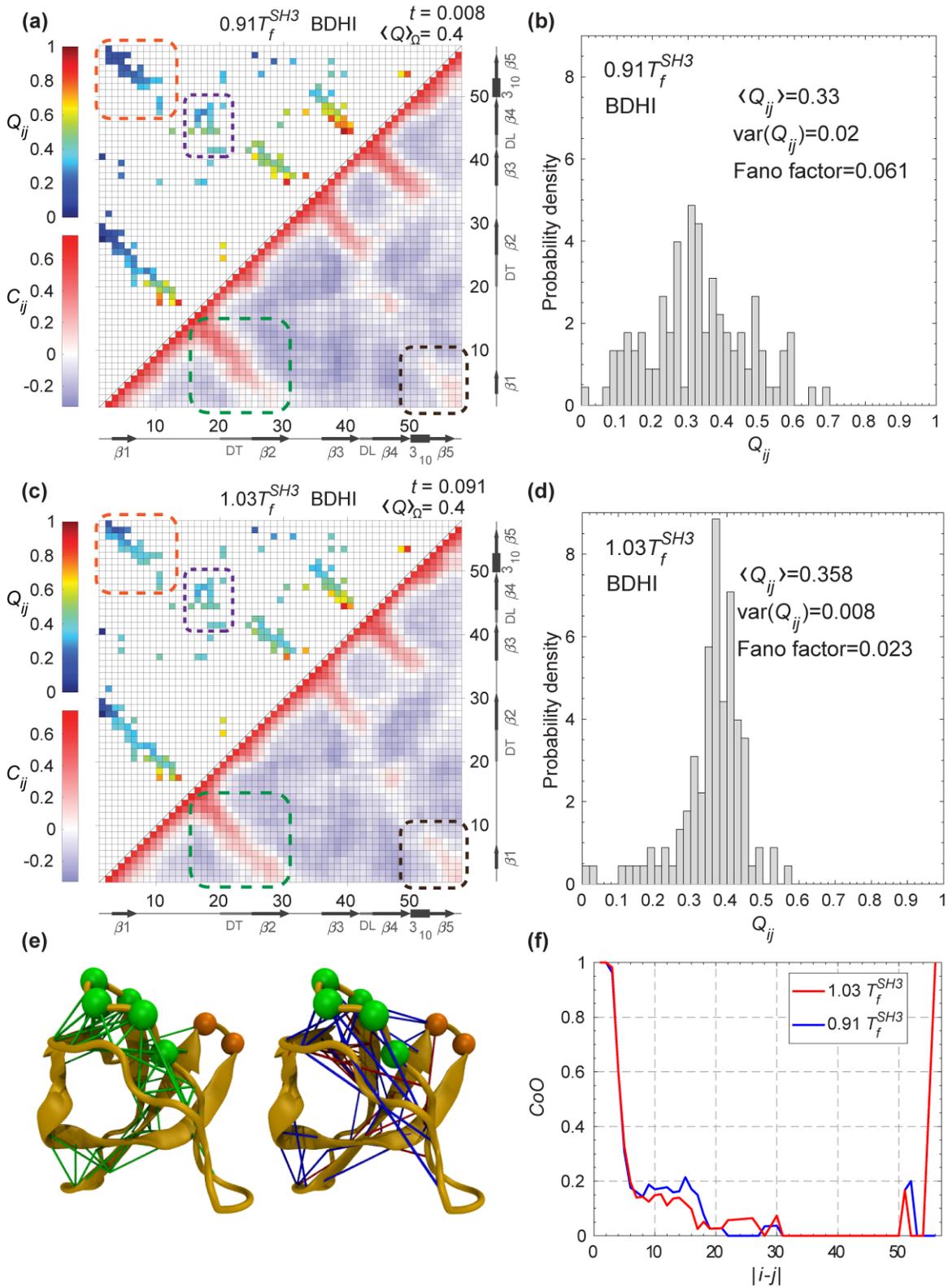

FIG. 8. (a) The probability of contact formation of each native pair $Q_{ij}(t)$ (upper triangle) and the displacement correlation $C_{ij}(t)$ (lower triangle) at $0.91T_f^{SH3}$ for SH3. (b) The distribution of native pairs from panel (a) with $|i$-$j|\geq$10. (c) $Q_{ij}(t)$ and $C_{ij}(t)$ at 1.03 $T_f^{SH3}$. (d) The distribution of native pairs from panel (c) with $|i$-$j|\geq$10. (a) and (c) correspond to a normalized time where $\langle Q \rangle_\Omega$ =0.4. The arrows and rectangles along (a) and (c) represent β-strands and helices, respectively. The orange and purple dashed boxes highlight the long-range contacts between the N-terminus and C-terminus, and the contacts between the diverging turn (DT) and the distal loop (DL), respectively. The black and green dashed boxes highlight the displacement correlation between the N-terminus and C-terminus, and between the N-terminus and both DT and β2, respectively. (e) The displacement correlation at 0.91 $T_f^{SH3}$ and 1.03 $T_f^{SH3}$ for all pairs are classified into three sets based on the magnitude of positive correlations. The coloring rules are the same as Fig. 7(e) with pairs with sequence separation greater than 7 residues and magnitude of displacement correlation above the threshold $\mu$ =0.095. (f) $CoO(|i-j|)$ for all pairs whose magnitude of the displacement correlation is above $\mu$ are organized according to the sequence separation $|i$-$j|$.

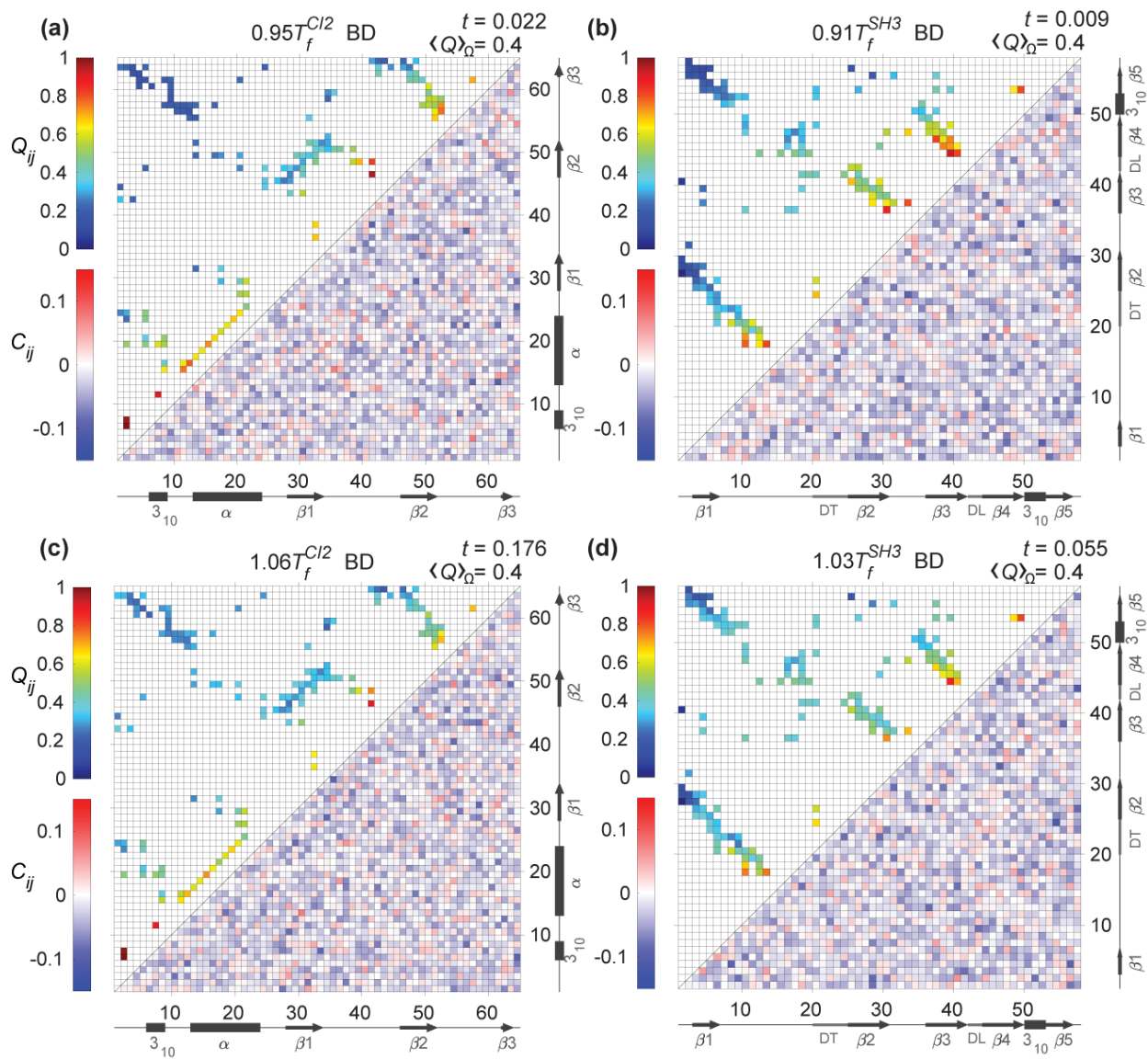

FIG. 9. The Brownian motion of residues without HI (BD) shows small and random displacement correlation for native and non-native pairs of CI2 and SH3. Upper and lower triangles represent the probability of contact formation for each native pair $Q_{ij}(t)$ and displacement correlation $C_{ij}(t)$, respectively. The panels are plotted at $T<T_f$ [(a) $0.95\,T_f^{CI2}$ and (b) $0.91\,T_f^{SH3}$ for CI2 and SH3, respectively], and $T>T_f$ [(c) $1.06\,T_f^{CI2}$ and (d) $1.03\,T_f^{SH3}$ for CI2 and SH3, respectively] at a normalized time where $\langle Q \rangle_\Omega = 0.4$. The arrows and rectangles represent β-strands and helices, respectively.

### E. HI can kinetically alter folding routes from multiple pathways

To further investigate the molecular underpinning of the crossover behavior that cannot be simply explained by the ratio of the effective diffusion coefficients from Sec. III B, we explored the possible changes in the pathways due to HI by projecting the kinetic trajectories on a two-dimensional free energy landscape. An additional reaction coordinate $Q_T$ involving a selected group of mid-range contacts from Figs. 7 and 8 (for CI2 and SH3, respectively), is employed to describe the folding process because we speculate the presence of hidden pathways that are not visible by a global parameter $Q$ [28].

#### 1.CI2

For CI2, $Q_T$ is defined as a set of the native contacts that are located in the neighborhood of the mini-core (contacts enclosed in the purple dashed rectangle of Fig. 7). Figure 10 reveals two distinct paths: one involves a high $Q_T$ (0.8) and other involves a low $Q_T$ (0.2) both at about $Q \approx 0.5$.

We projected two representative kinetic trajectories over the two-dimensional free energy surface as a function of $Q$ and $Q_T$. In Fig. 10(a), the kinetic trajectory, named route I, began from an unstructured chain. As time increases the α-helix and most of the contacts in $Q_T$ are formed before reaching $Q \approx 0.5$, which is the top of the barrier of the one-dimensional free energy profile as a function of $Q$. This involves the formation of the mini-core and contacts in the C-terminus before the formation of the hydrophobic core. After crossing the top of the barrier the hydrophobic core starts to form. Figure 10(b) illustrates another kinetic trajectory, called route II, started from another unfolded structure. As time increases, the contacts of $Q_T$ has not completely formed at $Q \approx 0.5$ while the hydrophobic core is formed before the mini-core. Then the contacts of the mini-core start to form to reach the folded state. Table III shows the number of trajectories that visit route I and II, and their corresponding average folding time. Route II is slower than route I for both BD and BDHI. BDHI accelerates the folding of both routes, and it reduces the number of trajectories that visit route II from 10.53% to 7.26%. BDHI not only reduces effective diffusivity, it also alters the folding route to favor a faster one than a slower one.

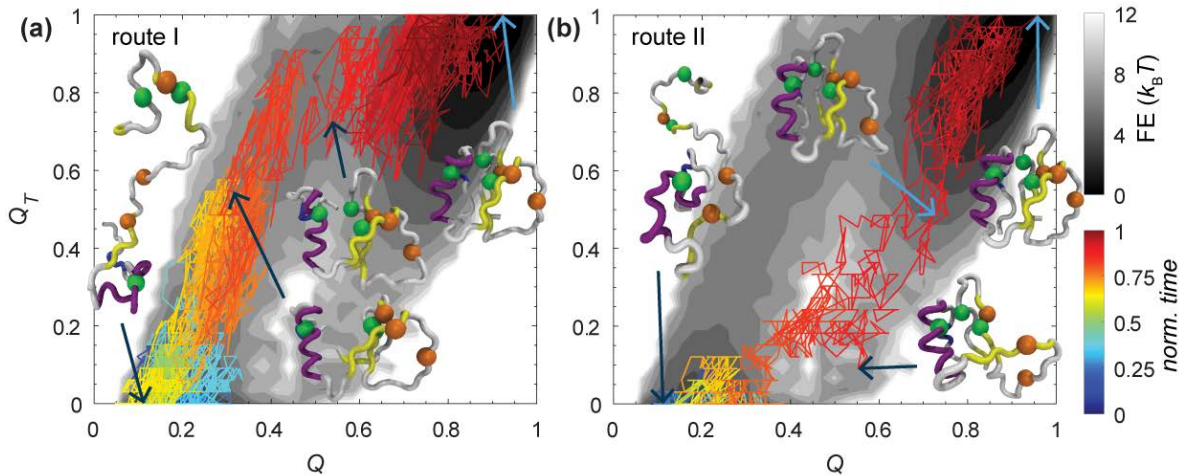

FIG. 10. Two representative kinetic pathways are projected on a two-dimensional free energy landscape of the fraction of native contact formation $Q$ and the fraction of native contact formation in the region of the mini-core $Q_T$ for CI2 at $0.95\,T_f^{CI2}$. Panel (a) shows a major pathway of folding kinetics (route I), representing a fast route, in the presence of HI. Panel (b) shows a minor pathway (route II). The folding free energy was colored in grayscale in units of $k_B T$. The kinetic trajectories were colored by normalized time (time divided by $t_{fold}$ of each trajectory) and projected on the folding free energy. Key conformations were selected for visual guidance. The significant residues for the hydrophobic core (Ala16, Leu49, and Ile57) and mini-core (Leu32, Val38, and Phe50) are illustrated with green and orange beads, respectively. The mini-core forms before the hydrophobic core in panel (a), whereas the opposite occurs in panel (b). Structures were created with VMD [30].

TABLE III. Number of trajectories and their folding time ($t_{fold}$) from a set of 1500 kinetic simulations that visit route I and II for CI2 at $0.95\,T_f^{CI2}$.

| | BD | | | BDHI | | $k_{BDHI}/k_{BD} =$ |
| Route | Number of trajectories | $t_{fold}^{BD}$ ($10^6\,\tau$) | | Number of trajectories | $t_{fold}^{BDHI}$ ($10^6\,\tau$) | $t_{fold}^{BD}/t_{fold}^{BDHI}$ |
|---|---|---|---|---|---|---|
| I | 1342 (89.47%) | 0.51 ± 0.01 | | 1391 (92.73%) | 0.38 ± 0.01 | 1.34 ± 0.04 |
| II | 158 (10.53%) | 0.76 ± 0.05 | | 109 (7.26%) | 0.51 ± 0.04 | 1.49 ± 0.15 |

## 2. SH3

As for SH3, $Q_T$ is defined as a set of the native contacts that are located in the neighborhood of the diverging turn (DT) and the distal loop (DL) (contacts enclosed in the purple dashed rectangle in Fig. 8). We created a two-dimensional free energy landscape as a function of $Q_T$ and $Q$ in Fig. 11. There is one dominant folding path. In fact, we checked whether a rare event second path occurs by raising the number of folding trajectories to 2500.

We projected a representative kinetic folding trajectory on the landscape. Figure 11 shows the pathway where the contacts between DL (orange beads) and DT (green beads) are formed before reaching $Q \approx 0.4$, which is the top of the barrier of the one-dimensional free energy as a function of $Q$. Then $Q_T$ increases along $Q$ and the rest of the protein forms to achieve the folded state. The formation of the contacts of DL and DT characterizes the selectivity of the transition state for SH3. This topological constraint may reduce the effect of HI on the folding kinetic rates at $T < T_f$.

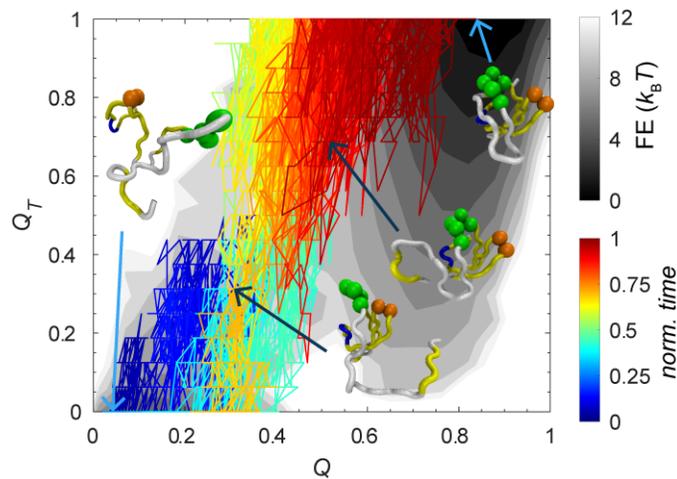

FIG. 11. One representative kinetic pathway is projected on a two-dimensional free energy landscape of the fraction of native contact formation $Q$ and the fraction of native contact formation in the region formed by the diverging turn (DT) and the distal loop (DL) $Q_T$ for SH3 at $0.91\,T_f^{SH3}$. It shows a dominant pathway of folding kinetics. The folding free energy was colored in grayscale in units of $k_B T$. The kinetic trajectory was colored by normalized time (time divided by $t_{fold}$ of the trajectory) and projected on the folding free energy. Key conformations were selected for visual guidance. The residues of the diverging turn (M20, K21, K22, G23, and D24) and distal loop (N42 and D43) are illustrated with green and orange beads, respectively. The native contacts between DT and DL are formed before reaching the top of the barrier. Structures were created with VMD [30].

## IV. DISCUSSION AND CONCLUSION

### A. Crossover behavior of folding kinetics on the non-Arrhenius curve

It has been shown extensively that protein-folding rates are temperature dependent. Folding rates with respect to temperature renders a U-shaped, non-Arrhenius curve where the rates are low at both low and high temperatures [43], and folding rates are fastest at a narrow range of temperature near $T_f$. Here, we show the additional impact of HI non-trivially affects folding in that it accelerates folding rates more than that of BD without HI at $T<T_f$. On the other hand, HI retards protein-folding rates more than that of BD without HI at $T>T_f$. To our knowledge, this crossover behavior of the folding times shown in Figs. 2(c) and 2(d) has never been observed or theoretically predicted. The temperature dependence of the effect of HI on folding and the crossover behavior might explain the mixed results of the HI influences on protein folding rates from several computational studies in the literature. In these previous studies, it was not clear whether the temperatures used are higher or lower than $T_f$. Rather, most simulation temperatures were justified by matching the experimentally measured diffusivity of a protein model. We will discuss previous work below.

Several groups used similar coarse-grained molecular simulations with a structure-based model for probing the impact of HI on protein folding. Their results vary: Kikuchi *et al.* [12] found that there is no clear difference in the folding kinetics with or without HI of a protein CI2 and two secondary structures, an α-helix and β-hairpin. Their $T_f$ values were not reported in their study, which makes it difficult to judge their results in an appropriate temperature regime where HI can accelerate or retard folding dynamics. Frembgen-Kesner and Elcock [10] studied 11 small proteins and also two secondary structures, an α-helix and β-hairpin. The folding time decreased with HI for all studied proteins, but it has the opposite effect for the secondary structures. It is inferred that they launched simulations at room temperature for all their systems. It may be that the folding temperatures of all proteins are greater than the room temperature used in the simulations, but it may not be the case for the secondary structures. Another study performed by Cieplak & Niewieczerzał [9] showed the folding time of three proteins (1CRN, 1BBA, and 1L2Y) over a range of temperatures. Although the differences of the folding time between their BD model with or without HI decreases at high temperature (above room temperature), there is no indication of a crossover behavior from their study. We speculate that their simulating temperatures are *not* close to $T_f$ because for a structure-based model that folds and unfolds in a two-state manner, the free energy barrier of protein folding is typically a few $k_B T$ at $T_f$; the folding time at $T_f$ is exponentially longer than the fastest folding time with a minimal free energy barrier. In addition, there is no clear evidence that the protein models remained thermally unfold at the maximum temperature studied with HI. Our work shows that the justification of the simulation temperature against the folding temperature is a criterion to assess the impact of HI on protein folding dynamics.

Additionally, Lipska *et al.* [11] argued that a structure-based model with only favorable attraction between native contacts is the reason why these studies mention above [9,10,12] have not observed retarded dynamics under HI from their simulations. They argued that the presence of intermediate states is key to a retarded dynamics by studying the effects of HI on two proteins (1BDD and 1EOL) at distinct temperatures with a coarse-grained molecular simulation using the UNRES force field. Indeed, HI can alter kinetic paths to favor non-productive intermediates at a temperature lower than the collapsed temperature $T_\theta$ as asserted by Tanaka [4]. Lipska's work has not investigate folding at $T>T_f$; thus, their conjecture is compatible with our work that HI can retard the folding dynamics at $T>T_f$.

### B. Underlying kinetic principles of the crossover behavior

The impact of HI on protein folding that gives rise to the crossover behavior is subtle over a wide range of temperatures because HI affects the folding mechanism in three thrusts that are not necessarily of equal prominence: (1) the dynamics of crossing over an activation barrier, (2) the choice of folding pathways, and (3) the motions between beads in viscous solvents. HI is a kinetic effect that expresses from a diffusion tensor in the equation of motion. It does not shape a folding energy landscape but it governs the ordering of contact pairs across a complex folding energy landscape particularly when more than one pathway from the unfolded state to the folded state exists. We argued that at $T<T_f$, the first two factors dominate the kinetic principle that HI accelerates folding creating a backflow. In addition, the extent of acceleration is dictated by a protein's topology. At $T > T_f$, the third factor from non-native contacts arises and HI retards folding.

First, the development of an appropriate reaction coordinate that best describes the profile of a free energy barrier often relies on the characteristics of the energy landscape [51]. For a minimally frustrated energy landscape that resembles a funnel [16], it has been shown by the use of structure-based models, like the one we employed, that the fraction of the native contact formation $Q$ is a reasonably "good" 1-D reaction coordinate [26] to predict protein folding rates simply from the features of an activation barrier and the shape of the unfolded basin [43]. As a protein model becomes complex, these features become less harmonic and the diffusivity of $Q$, $D(Q)$, becomes dependent on $Q$ itself. Wang *et al*. [28] have shown that prediction of rates improves by including a $Q$-dependent diffusivity. However, their procedure to compute $D(Q)$ is to constrain selected structures that diffuse with a specific $Q$ under a harmonic potential well. Whitford *et al*. showed that the computation of an effective diffusivity $D^{eff}$ is sufficient for the computation of rates when it is fitted from a reasonably linear region of a mean-square displacement versus time plot [42] for a specific site-tRNA movement in a ribosome. In our study, this projection was performed onto a single reaction coordinate, the fraction of all native contact formation $Q$, to obtain the one-dimensional free energy landscape. In a qualitative sense the diffusion coefficient of $Q$ can describe the ratio of the folding rates between BD and BDHI at $T < T_f$ and $T > T_f$ for CI2 and SH3. However, the ratio of the folding rates in the presence or absence of HI *cannot* be fully explained by the ratio of effective diffusion coefficients (Table I and II). Even when we computed $D^{eff}$ from a selective contact pairs that are pertinent in the formation of the transition states ($Q_T$), $D_{BDHI}^{eff} / D_{BD}^{eff}$ based on $Q_T$ (1.21 ± 0.00 for CI2 and 1.10 ± 0.00 for SH3) becomes closer to $k_{BDHI} / k_{BD}$ than the values from Table I and II at $T < T_f$. However, at $T > T_f$, it becomes much worse (0.11 ± 0.00 for CI2 and 0.22 ± 0.00 for SH3). It is indicative that there exists at least two competing mechanisms requiring more than one reaction coordinate to fully describe an energy landscape as discussed by Yang and Gruebele [52]. They showed at least two reaction coordinates that are opposing to one another are required for describing folding of a small protein over a full temperature range by mutagenesis. It supports our speculation that the search of a single perfect order parameter may not fully solve the mystery of the crossover behavior under HI.

Second, since folding is a complex process (NP-complete [53,54]) moving on a high dimensional energy surface, we lose information when projecting the landscape onto a single order parameter. We argued that a single reaction coordinate, no matter how optimally it is defined, will still fall short of addressing a folding mechanism that shows competing folding pathways. As shown by the use of $Q_T$, the folding energy landscape is extended into another dimension to capture the folding kinetics. For CI2, there are two distinct pathways. One pathway forms the mini-core first, while the other pathway forms the hydrophobic core first. At $T<T_f$, HI will change the kinetic pathway that favors the formation of the mini-core (route I), which is the faster of the two pathways. In the presence of HI, there is a flow of configurational space towards the folded state that increases the number of trajectories through route I. This, in turn, increases the folding rate. HI may guide the folding process and prevent kinetic traps from less populated routes. The presence of a "hidden" pathway for CI2 has also been addressed by Wang *et al.* as a justification of using kinetically determined variables to calibrate the equations from the energy landscape theory to predict folding rates [28]. The Weeks group used a simple HP model of a protein, consisting of only hydrophobic (H) and hydrophilic (P) spherical monomers to show two reaction coordinates with distinct diffusion coefficients necessary for describing a meaningful collapse mechanism [55]. As for SH3, folding follows a specific nucleation site as an obligatory step for folding. It folds through the formation of high ϕ-value amino acids (DT and DL) [34,35]. We argued that because SH3 has only one dominant pathway, HI has less of an effect on the folding rate as that of CI2 at $T<T_f$. Additionally, several single molecule pulling experiments have also shown that the addition of more than one order parameter is necessary to describe folding [56,57]. Even though Woodside and coworkers have shown that it is possible to use a single reaction coordinate to characterize the entire folding landscape for a *specific protein,* they are not able to claim that it is possible for *all* proteins, however [58]. In general, a multi-dimensional projection is necessary to capture all the information of the energy landscape.

Lastly, at $T>T_f$, protein polymers are in a good solvent where the beads favor interaction with solvents; thus a protein model unfolds. Without a defined folding nucleus, the dispersion of the probability of contact formation among native contacts becomes narrow. In other words, the probability of contact formation between long-range contacts [Figs. 7(c) and 8(c), upper triangle] is more probable at $T > T_f$ than at $T < T_f$. For BDHI, we noticed a strong displacement correlation between non-native pairs at the proximity of native pairs [Figs. 7(c) and 8(c), lower triangle]. Under HI at $T > T_f$, a protein takes more time to fold, reflecting excess hydrodynamic friction for fluid drainage due to the close proximity between patches of native and neighboring non-native contacts separated at a long distance in a sequence space [Figs. 7(f) and 8(f)].

## C. Possible experimental validation

Can we experimentally determine the effect of HI on protein folding? Or is there evidence to support our investigation? We argue that the experiment of temperature dependent mutagenesis of protein folding ($\phi$-value experiments) by the Gruebele group may provide experimental validation of our predictions. Experimentalists mutate amino acids on a protein one by one and then measure the change in the folding rates. With this difference compared against the change in protein stability, experimentalists map out the information of the transition state of protein folding from CI2 decades ago [59]. Normally it is experimentally determined at a room temperature. There can be a $\phi$-value for each amino acid that ranges from 0 to 1. A $\phi$-value at mid-range denotes the importance of that residue forming transition states. However, Gruebele's group has measured the $\phi$-values for a simple $\lambda$ repressor over a full range of temperatures, which is a notable departure from the single temperature measurements [52]. As expected, they noticed a quadratic curve of the logarithm of folding rates with respect to inverse temperature (shown in Fig. 3 of [52]). Most of his mutations were done for amino acids at the hydrophobic core such that the mutation can impact both folding kinetics and stability. We noticed, however, that a few mutations (e.g., $\lambda$sQ33Y, $\lambda$sA37G, and $\lambda$sA81G) close to the loop or the end of the protein show an interesting behavior: they fold faster than the wild-type (WT) at low temperature and become slower than the WT at a high temperature. We speculate this can be a signature of the impact of HI since these mutations do not directly affect the protein's thermodynamic stability, as these mutations are not in the hydrophobic core of the protein. To show a full impact from HI on folding, mutations should be performed on the loop regions that affect the kinetics but not necessarily the stability of folding. We predict that the outer loop mutations will show a crossover behavior at a full range of temperatures when compared to the wild-type protein due to the impact of HI and not because of thermodynamic stability.

## D. Concluding Remarks

In summary, we have settled the controversy of the extent of the impact of HI on folding kinetics by comparing the effect of HI over a range of temperatures instead of a single temperature. We found that HI can both accelerate protein folding at a temperature lower than the folding temperature and retard protein folding at a temperature higher than the folding temperature, in comparison with the folding dynamics without HI. Through this result, we have explored three different causal mechanisms: (1) the effective diffusive dynamics of $Q$ over the free energy barrier crossing, (2) multi-dimensional landscape gives rise to a choice of multiple folding pathways, and (3) the kinetic ordering and hydrodynamically correlated motion between beads in viscous solvents. Finally, we have proposed experiments to test our predicted results. Our findings will provide theoretical insight to future protein folding kinetic experiments, and guide simulation design for coarse-grained models.

## ACKNOWLEDGMENTS

We thank the members from the Center for Theoretical Biological Physics (CTBP) at Rice University and Dr. Greg Morrison for stimulating discussions. MSC thanks Dr. Jeff Skolnick for the software HIBD. We thank Basilio C. Huaman and Mohammadmehdi Ezzatabadipour for their early participation in this project. We thank the Center for Advanced Computing and Data Science and Research Computing Center at the University of Houston for the computational resources. We thank the National Science Foundation for their funding support (MCB: 1412532, ACI: 1531814, PHY: 1427654).